# Phase Transitions of Civil Unrest across Countries and Time

Dan Braha[1, 2, *]

[1] University of Massachusetts, Dartmouth, MA, USA and

[2] New England Complex Systems Institute, Cambridge, MA, USA

* Email: braha@necsi.edu

Phase transitions, characterized by abrupt shifts between macroscopic patterns of organization, are ubiquitous in complex systems. Despite considerable research in the physical and natural sciences, the empirical study of this phenomenon in societal systems is relatively underdeveloped. The goal of this study is to explore whether the dynamics of collective civil unrest can be plausibly characterized as a sequence of recurrent phase shifts, with each phase having measurable and identifiable latent characteristics. Building on previous efforts to characterize civil unrest as a self-organized critical system, we introduce a macro-level statistical model of civil unrest and evaluate its plausibility using a comprehensive dataset of civil unrest events in 170 countries from 1946 to 2017. Our findings demonstrate that the macro-level phase model effectively captures the characteristics of civil unrest data from diverse countries globally and that universal mechanisms may underlie certain aspects of the dynamics of civil unrest. We also introduce a scale to quantify a country's long-term unrest per unit of time and show that civil unrest events tend to cluster geographically, with the magnitude of civil unrest concentrated in specific regions. Our approach has the potential to identify and measure phase transitions in various collective human phenomena beyond civil unrest, contributing to a better understanding of complex social systems.

# INTRODUCTION

Phase transitions, which are marked by abrupt changes in the features of a system, occur in both nature and society (1, 2). Understanding the transitions between disordered and ordered system structures is a prevalent theme underlying research in this field (3). Over the past few decades, there has been a growing trend towards utilizing statistical physics (4), biology (3), and ecology (5) to examine the dynamics of social phenomena (6—10). These interdisciplinary approaches have proven to be particularly useful in understanding phase transitions within social systems. As a result, significant attention has been devoted to investigating phase transitions exhibited by models that attempt to capture the essential characteristics of specific social phenomena. The demonstrability of phase transitions in stylized models of social systems is wide in scope (6), encompassing diverse fields such as traffic flow (11), agent-based models in econophysics (10, 12), evolutionary game theory (8), opinion dynamics (7, 13), cultural dynamics (7, 14), crowd behavior (7), language dynamics (7), criminology (6), and social collapse (15), to name just a few examples.

Recent studies (6, 7, 16) have drawn attention to a gap between the theoretical modeling of phase transitions in social systems and their empirical verification. While empirical research on this topic is limited, self-organization and phase transitions have been proposed in various domains, including human dynamics (7, 17, 18), financial markets (10, 19—22), experimental econophysics (23), opinion dynamics (16), and elections and voting behavior (6, 24). Noteworthy to our investigation are empirical findings related to patterns of protest recruitment bursts on the Twitter network (25), oscillating patterns of political instability (26, 27), and the use of Markov models (28—30) to forecast different phases of armed conflict dynamics (31). Although the empirical exploration of phase transitions in social systems remains scarce, these examples underscore the potential for further investigation and the necessity for greater emphasis on empirical validation.

To fill the aforementioned gap, this study outlines statistical methods designed to detect and assess the plausibility of phase transitions in large scale social systems. These methods are specifically applied to analyze civil unrest, one of the most potent forms of collective human dynamics. Our analysis utilizes a comprehensive dataset spanning 170 nations from 1946 to 2017. The results shed light on the seemingly unpredictable onset of civil upheaval, supporting the idea that large scale collective civil unrest unfolds in distinct recurring phase transitions that can be precisely identified and measured. This finding aligns with prior efforts in modeling civil unrest as a self-organized process (see Supplementary Methods) and establishes a link between phase shifts in civil unrest and other collective phenomena observed in complex systems.

Civil unrest is a ubiquitous collective social phenomenon that has affected every aspect of social life throughout history, including human rights (32—36), economic issues (37—46), independence movements (47—50), and religion (51). Civil unrest is characterized by a sudden and spontaneous eruption, resulting from an unintended chain of historical events, which leads to a significant disruption of regional and global order, often followed by periods of relative calm. Moreover, the interdependence among various parts of society through communication and social networks facilitates the spatial and temporal coordination of protest, riot, and rebellion events across diverse communities, regions, and even countries (52). Several prominent instances of such widespread collective behavior comprise the African American, Hispanic, and Latino civil rights movement during the 1950s and 1960s (53), the series of political protests and revolutions that swept across northern Africa and the Middle East in 2010 and 2011 (54), the worldwide demonstrations against the multinational corporation Monsanto held in 436 cities across 52 countries (41), the 2011 Occupy Wall Street movement against economic inequality that originated in New York City and spread to cities worldwide (37), and the independence movements of Eastern European nations in the late 1980s that resulted in the dissolution of the Soviet Union in 1991 (47).

Although there has been a perceived rise in the frequency and intensity of anti-government demonstrations over the past decade, with the GDELT Project reporting an average annual increase of 5 to 19 percent across different regions of the world (55), there is no consistent pattern of disorder in terms of the number of people expressing dissatisfaction with government policies or authority, the level of violence in protests, the number of casualties, or the extent of property damage (52, 56—58). Civil unrest in various countries worldwide, like other collective phenomena in complex systems such as earthquakes (59), financial crises (20), traffic jams (11), and social collapse (15, 27), is characterized by extended periods of relative calm that are punctuated by sudden bursts or waves of high-intensity, large-scale civil disorders involving numerous participants and violent events (56, 60—63). The bursty behavior observed during instances of civil unrest can be attributed, in part, to the contagion and social influence mechanisms that operate among individuals connected by communication and social networks across different regions (25, 52, 64—75). A close visual examination of civil unrest activity across different parts of the world suggests a considerable association between a country's level of civil disturbance and that of its neighboring countries (see Results). This phenomenon of geographic clustering of civil unrest can be attributed, at least in part, to the mutual influence that neighboring countries have on each other's civil unrest levels, which could potentially facilitate the spread of unrest across international borders (52).

The approach outlined in this paper for detecting regime changes of civil unrest remains viable, even when a comprehensive understanding of all the mechanisms driving civil unrest is incomplete. This is especially pertinent given the difficulties in constructing precise mechanistic models for human societies. This approach is in accordance with related research that emphasizes specific early-warning signals preceding significant shifts in complex systems (1, 2, 20), even when facing challenges in constructing precise mechanistic models for these intricate systems. While constructing detailed quantitative mechanistic models for various complex systems, including civil unrest, poses a significant challenge, formulating plausible and realistic models that elucidate observed patterns in real-world systems and anchor empirical research in theoretical foundations substantially enhances the credibility of hypotheses.

The intermittent and complex nature of civil unrest could be accounted for by alternative mechanistic models. One successful class of mechanistic models for explaining the dynamics of a wide variety of complex systems, including civil unrest, is self-organized criticality (52, 58, 76—79). Such models describe the dynamics of complex systems, including forest fires, earthquakes, and epidemics, as being governed by a slow driving force, either physical or informational. This driving force gradually pushes the system into a highly vulnerable state, leading to the dissipation of energy in avalanches through the system's spatial degrees of freedom (80—82). For instance, the mechanistic micro-dynamic model of civil unrest in (52), which is an expanded version of the forest-fire model (81) of self-organized criticality, can effectively replicate the dynamics observed in real-world civil unrest, particularly the transitions between periods of calm and intense civil unrest. The model, as detailed in Supplementary Methods, operates on a spatial network that connects different regions within a country. Over time, social, economic, and political stress builds up in geographic sites, making them susceptible to unrest with an 'unrest susceptibility' probability. Spontaneous outbursts of social unrest may occur in susceptible regions with a 'spontaneous outburst' probability, which can quickly spread to nearby and distant vulnerable sites with an 'infectiousness' probability. Such activity can trigger instabilities within the network and cause avalanches of disturbances that percolate throughout the country, leading to a cascading effect of further unrest. Simulating the micro-dynamics mechanism described above has demonstrated that this approach can effectively account for the observed phenomenon where significant unrest events disrupt a calm state at irregular intervals (see Supplementary Figure 1). Moreover, the simulation faithfully reproduces both the scale and frequency of riots and protests witnessed globally over nearly a century (52).

According to the above micro-dynamic civil unrest model (52), the spatial interactions among a country's elements, including people or regions, can have an intriguing implication: they can predict the presence of a narrow transition region that separates two distinct macroscopic phases of social instability (see Supplementary Figure 1). These phases encompass low- and high-intensity unrest and turmoil events, and the identification of such a transition region can be an indicator of the likelihood and severity of social instability within a country. The two-phase behavior of civil unrest can be further explained by considering the time scale separation that often characterizes riots, unrest, and revolutions (52, 83, 84). That is, the rate at which unrest propagates from disrupted regions to adjacent susceptible areas is significantly faster than the rate at which susceptible regions become vulnerable to unrest due to social, economic, and political stress. Additionally, the rate at which civil unrest is spontaneously triggered in susceptible regions is markedly slower than the first two rates. When the density of susceptible regions falls below a critical threshold (a low-intensity latent phase), civil disturbance is unlikely to spread, and collective disorder will quickly dissipate, even in the presence of spontaneous triggering events and high infectiousness rates between the nearest and most distant susceptible geographic regions. However, once the critical threshold is exceeded (a high-intensity latent phase), a vast cluster of connected regions forms, and spontaneous hostile outbursts would likely spread to many parts of society, with the severity of disorder swiftly increasing and then subsiding. When new susceptible regions emerge, this pattern tends to repeat itself, with social tensions building up within the society. If the critical threshold is exceeded again, collective disorder may ensue, perpetuating the cycle (see Supplementary Methods).

The micro-model discussed above proposes a phenomenological perspective on civil unrest, framing it as a collective phenomenon undergoing self-organization and giving rise to recurring phase transitions. The initial phase is characterized by a gradual shift in societal attitudes towards hostility against the established social order. This phase, which can last for several years, is marked by the accumulation of social, economic, and political stress in certain segments of the population, making them more susceptible to different forms of collective unrest. The second phase typically begins with one or more triggering events, followed by mobilization, mass formation, and the emergence of widespread protests and civil unrest. During this second phase, communication channels, such as rumors, social networks, and, more recently, social media, help disseminate shared grievances among various communities, inciting further mobilization and expanding the intensity and scope of civil disorder.

It is crucial to emphasize that observing civil unrest following a discontinuous phased pattern could be explained by various alternative mechanisms and does not definitively confirm self-organized criticality; it merely posits it as a plausible explanation consistent with empirical research. Rather than presenting a precise representation of civil unrest mechanisms, models of this nature simulate the gradual accumulation of tensions (or vulnerabilities) during a relatively calm state, eventually released through sudden transitions into widespread civil unrest. Given the intricate dynamics of civil unrest, we regard the presented mechanistic model as a phenomenological representation, offering plausible insights into the potential for recurrent critical transitions in real-world civil unrest data—an aspect that necessitates empirical investigation. The latter is the focus of this paper.

This paper aims to investigate the hypothesis of repeated latent phase shifts in civil unrest across nations and time periods, by addressing several questions: Is it plausible to characterize the dynamics of collective civil unrest as a sequence of repeated phase shifts, with each phase displaying identifiable and measurable latent characteristics? Does the hypothesis of phase shifts imply that certain aspects of civil unrest are controlled by universal mechanisms, despite the unique characteristics of individual countries and geographic regions? To what extent do countries worldwide share similarities or disparities in the intensity of civil disturbance? Moreover, the paper aims to explore the relationship between geographical embeddedness and the intensity of a nation's long-term civil unrest per unit of time. To accomplish these objectives, the paper employs empirical data from various nations and time periods.

It is important to note that the approach proposed in this paper for investigating the discrete latent phases of civil unrest differs considerably from the typical descriptive or ad hoc references to "waves," "spikes," and "bursts" of civil unrest (56, 60—63, 85). Our approach entails identifying *latent phases* that are inherent in the dynamics of civil unrest and can be accurately measured and detected. Importantly, each latent phase type represents the same underlying phenomenon, regardless of when it occurs, and transitions between latent phases are characterized by well-defined probabilities. To illustrate this concept, we draw an analogy between "latent phases" and "climate," and "bursts" and "weather." Whereas weather refers to rapidly changing short-term atmospheric conditions, climate describes long-term patterns and trends in those conditions over time. Similarly, a seemingly significant burst of civil unrest may occur during a low-intensity phase, while a string of peaceful activities may occur during a high-intensity period. The methods put forth in this paper present a rigorous and systematic way of detecting and analyzing these distinct latent phases of civil unrest, potentially leading to a deeper understanding of the underlying mechanisms behind civil disturbance.

In this study, we examine civil unrest incidents reported in newspapers (86) from 170 countries spanning the years 1946 to 2017 (see Methods). To address our research questions, we employ a macro-level statistical model of civil unrest that leverages a statistical hidden Markov model (87—96) to test the central hypothesis of whether recurring latent phases underlie civil unrest events across countries and time. Our study provides compelling evidence supporting the central hypothesis of this paper, demonstrating the existence of quantifiable latent phases of civil unrest in the vast majority of countries. Using our macro-level phase model, we can accurately gauge the expected magnitude and variability of civil unrest across different latent phases in each country, as well as the likelihood of transitioning between these latent phases. Moreover, the macro-level phase model enables us to assess the long-term proportion of time that a country spends in low- or high-intensity civil unrest phases, as well as its long-run magnitude scale of civil unrest, which extends beyond the noise inherent in the data. Our findings also highlight the worldwide implications of civil unrest that transcend national boundaries. This is demonstrated by the identification of regional clustering of civil unrest events across the world. While this phenomenon does not necessarily imply causation, it suggests that the tendency for civil unrest in one nation to influence those in neighboring countries may partially explain the observed clustering effect.

## RESULTS

**Overview of methodology**

We propose a macro-level phase model, which is a statistical Markov model (see Methods), to describe the macroscopic behavior of civil unrest. Similar methods have been applied in a variety of fields, including bioinformatics (93), geophysics (94), finance (95), ecology (96), and armed conflicts (28—30). In this study, we utilize the statistical Markov model not as a predictive tool (though see its potential for forecasting in Discussion and Future

Work), but as a rigorous means to assess the viability of our central hypothesis using a Monte Carlo Kolmogorov-Smirnov two-sample testing (see Methods).

Specifically, we investigate the presence of recurring latent phases that underlie civil unrest events throughout different countries and historical periods. Our analysis encompasses a wide range of events, including assassinations, general strikes, guerrilla warfare, government crises, purges, riots, revolutions, and anti-government demonstrations, which are incorporated in a comprehensive long-term event dataset from 1946 to 2017 for 170 countries (see Methods). For each country, the various civil unrest events are aggregated, and a 'weighted conflict metric' time series is obtained. The magnitudes of civil unrest were then calculated by taking the logarithm of the weighted conflict metric. The macro-level phase model (outlined in Methods) postulates, for any particular country, that the distribution generating the observed level of civil unrest magnitude at any given time is a function of the current latent phase of an unobserved Markov process. The macro-level phase model consists of two components: an unobserved $m$-state Markov chain that governs the dynamics of the underlying latent phases of civil unrest, and a phase-dependent process that characterizes the probability density functions generating the observable civil unrest magnitudes at any given time. The latter process is contingent on the current latent phase of civil unrest. More specifically, the macro-level phase model posits that the magnitude of civil unrest in any given year is drawn from one of $m$ normal distributions. The normal distributions chosen are dependent on the current latent phase of the underlying Markov chain, which governs the transitions between the $m$ latent phases. By employing phase-dependent normal distributions to represent the magnitudes of civil unrest, we suggest that the associated weighted conflict metrics adhere to lognormal distributions. This approach enables the development of a flexible macro-level phase model that effectively captures both heavy-tailed and thin-tailed distributions of civil unrest events (52). Our proposed model involves $m(m+2)$ parameters and is estimated by maximizing the likelihood of the observed civil unrest magnitude data (see Methods). To determine the appropriate number of latent phases $m$ for each country, we estimated models with varying numbers of latent phases and applied the Bayesian information criterion (BIC) (see Methods) to determine the appropriate number of latent phases. Once a hypothesized macro-level phase model has been selected, we assess the model's suitability to the observed unrest dataset using a range of goodness-of-fit Monte Carlo tests (see Methods). The overall methodology is summarized in Figure 1.

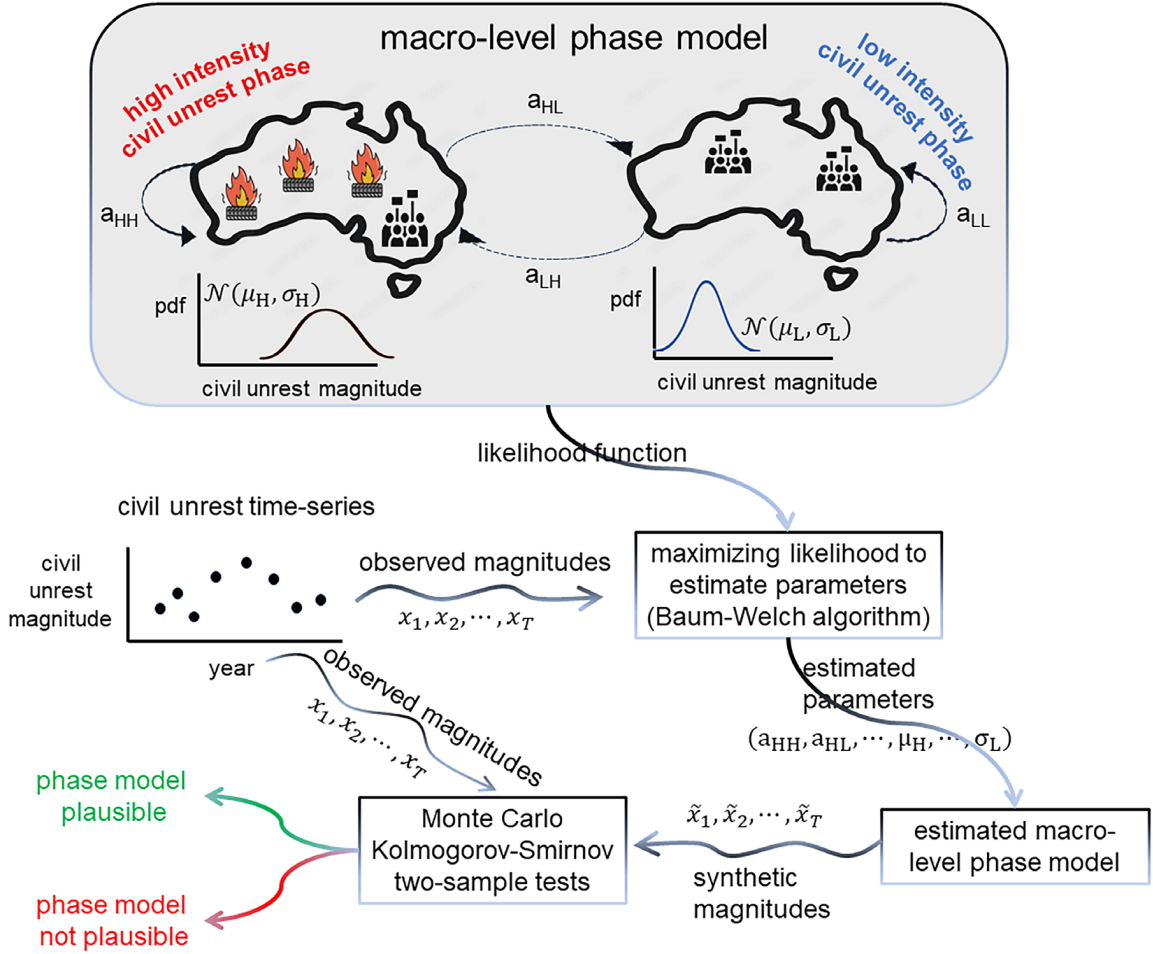

**Fig. 1. Overview of methodology**. The research methodology proposes a macro-level model governing the dynamics of unobserved latent phases of civil unrest, inspired by self-organized micro-level mechanisms of civil disorder (52, 58, 76—79; see example in Supplementary Methods). Using 'weighted conflict metric' time series data from 1946 to 2017 for 170 countries, we define civil unrest magnitude as the logarithm of this metric. Subsequently, we utilize the Baum-Welch algorithm to estimate parameters of the macro-level phase model, such as transition probabilities and phase-dependent probability distributions of civil unrest magnitudes. The algorithm calculates parameter values that maximize the likelihood of the observed civil unrest magnitude data. Detailed estimates of these parameters for each country are presented in Supplementary Figures and Tables. To validate the model against the unrest dataset, we conduct various goodness-of-fit Monte Carlo tests. The parameter $a_{HL}$ represents the transition probability from the high-intensity to the low-intensity phase. Parameters $\mu_H$ and $\sigma_H$ denote the unobserved mean and standard deviation respectively, associated with the phase-dependent normal probability density function of civil

unrest magnitudes during the high-intensity latent phase H. Other parameters are defined in a similar fashion. The vector $(x_1, x_2, \cdots, x_T)$ represents actual civil unrest magnitude values collected over $T$ periods (years). Correspondingly, the vector $(\tilde{x}_1, \tilde{x}_2, \cdots, \tilde{x}_T)$ represents synthetic data generated by the macro-level phase model. Further elaboration on this methodology is provided in the Methods section.

**Assessing the plausibility of the recurring phase shifts hypothesis**

Figure 2 illustrates the results of the macro-level phase model estimated on civil unrest time series for a representative sample of countries (for comprehensive results, refer to Supplementary Figures and Tables). The model discerns distinct latent phases of civil unrest across countries, with 81 countries exhibiting two latent phases labeled "low" and "high" (see Methods), five countries showing three latent phases labeled "low," "intermediate," and "high," while 64 countries demonstrate a single arbitrary phase. A single latent phase represents a phase model where the magnitude of unrest in a given year is drawn independently from a normal distribution, or equivalently, the weighted civil conflict metrics are drawn from a lognormal distribution.

To illustrate, consider the dynamics of civil unrest in Spain, which can be classified into two distinct latent phases. The first latent phase is marked by high-intensity civil unrest, while the second latent phase is associated with lower-intensity unrest. During the high-intensity latent phase, the magnitude of civil unrest is distributed normally with a mean of 8.37 and a standard deviation of 0.35, whereas in the low-intensity latent phase, the unrest magnitudes follow a normal distribution with a mean of 6.84 and a standard deviation of 0.96. There is a 94% probability that another high-intensity latent phase will follow a year characterized by a high-intensity phase. Following a low-intensity latent phase, there is an 87% likelihood that the following year will also be a low-intensity latent phase. By combining these probabilities with the corresponding phase-dependent normal distributions specified above, an alternating sequence of latent phases can generate observable magnitudes of civil unrest. This characterization of civil unrest draws attention to the distinction between latent phases and observable phenomena such as waves, spikes, or bursts of civil unrest activity. For instance, during a low-intensity latent phase, an observed activity with a magnitude of 8.76, which is two standard deviations above the mean, may seem like a sudden spike or wave of civil unrest activity. However, it could also be a random fluctuation, underscoring the importance of exercising caution when interpreting observed magnitudes of civil unrest. This insight provides a more nuanced understanding of the complex dynamics of civil unrest and highlights the need for robust statistical methods, such as the macro-level phase model, for analyzing and interpreting civil unrest data.

Figure 3 presents the results of the goodness-of-fit tests conducted on a representative sample of countries to evaluate the plausibility of the postulated macro-level phase model as a representation of the unrest data (for comprehensive results, see Supplementary Figures and Tables). The macro-level phase model effectively captures crucial aspects of the civil unrest magnitude data. Specifically, it accurately reproduces the marginal distribution of civil unrest magnitudes (the marginal distribution is defined in Methods). Moreover, it successfully captures key summary statistics such as the mean, median, standard deviation, lower and upper quartiles, and minimum and maximum values of the civil unrest magnitude, highlighting its ability to represent the overall behavior of the data. The goodness-of-fit tests indicate that the hypothesis of recurring latent phases, as a fundamental feature of collective civil unrest dynamics, is nicely supported by real-world civil unrest data from a diverse range of countries. These results suggest that the macro-level phase model is a credible framework for understanding civil unrest dynamics in various contexts.

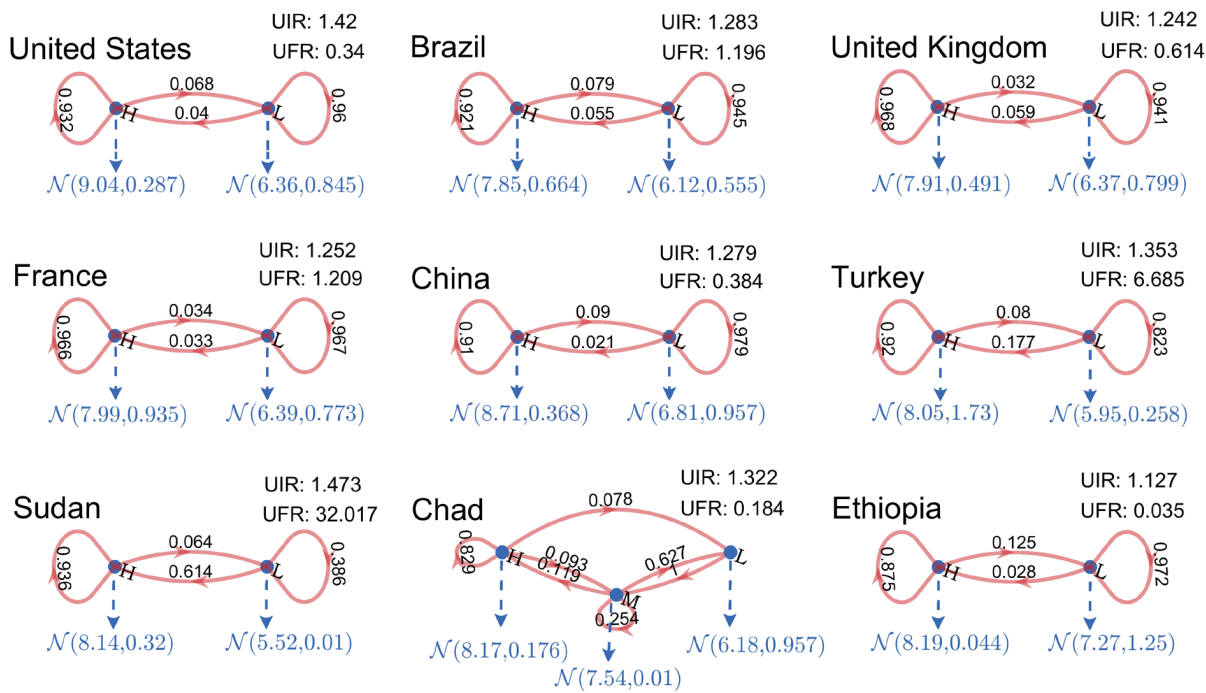

**Fig. 2. Directed graphs of the macro-level phase model for a sample of countries.** The unobserved low- intermediate- and high-intensity latent phases are denoted by 'L,' 'M' and 'H', respectively. Each directed graph shows the transition probabilities of the underlying Markov chain and the parameters of the phase-dependent normal distributions. Here, we label the latent phases (high, intermediate, or low) according to the mean values of the corresponding normal distributions. The Unrest Intensity Ratio (UIR) is defined as $\mu_H/\mu_L$, and the Unrest Fluctuation Ratio (UFR) is defined as $\sigma_H^2/\sigma_L^2$. Supplementary Figures and Tables contains the estimated parameters of the macro-level phase model for all countries.

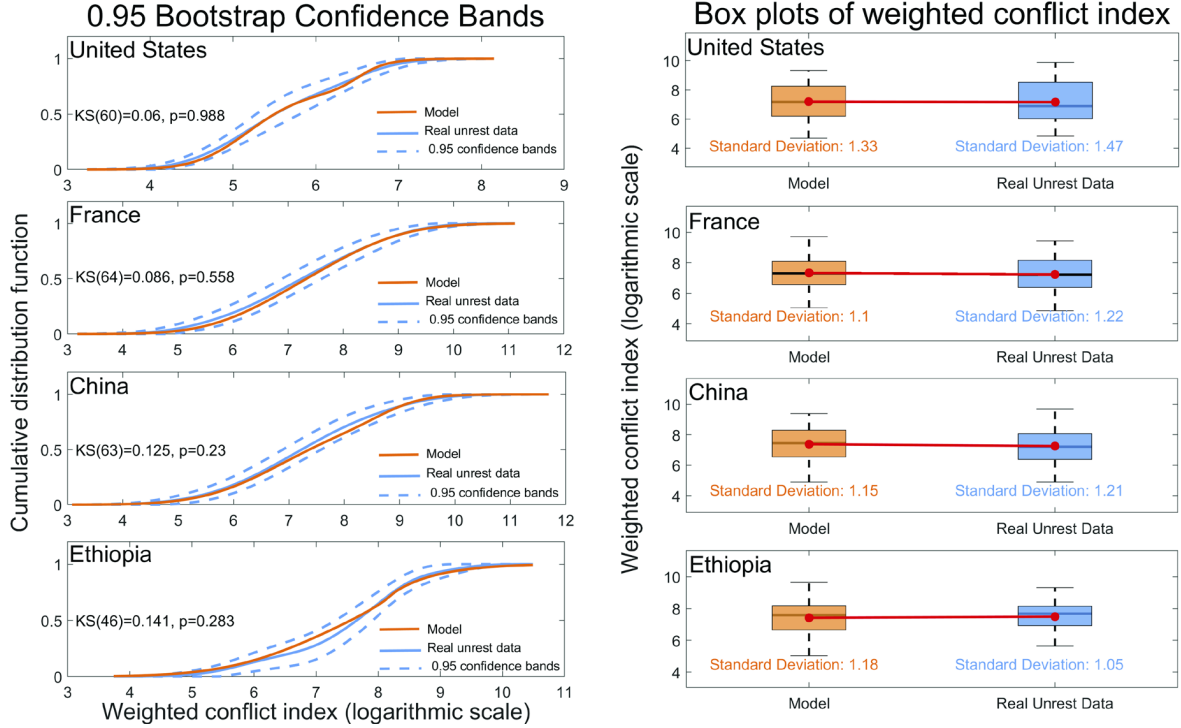

**Fig 3. Goodness-of-fit tests of the macro-level phase model.** The left panels display the cumulative distribution functions (CDFs) of civil unrest magnitudes generated by the macro-level phase model (orange solid line) alongside the CDFs of the empirical civil unrest data (blue solid line), for a representative sample of countries (see Supplementary Figures and Tables for a complete list of countries). Additionally, 95% confidence bands (blue dashed lines) are shown for kernel density estimates of the empirical civil unrest data, which were generated using the bootstrap method averaged over 1000 replications. For each country, we report the Kolmogorov-Smirnov (KS) statistics obtained through a Monte Carlo procedure, as well as the corresponding p-values for the fit to the macro-level phase model (see Methods). Our analysis reveals that the p-values for all distributions are greater than the α=0.05 significance level threshold, indicating that the macro-level phase model is a plausible hypothesis for the civil unrest data. Since our statistical hypothesis testing is performed at the individual country level and does not involve controlling for multiple comparisons, there is no necessity to utilize the Bonferroni-adjusted alpha. The right panels of Figure 3 display box plots that compare the empirical civil unrest distributions with the marginal distributions of civil unrest magnitudes generated by the macro-level phase model. The outer edges of the box represent the first quartile Q1 (the 25th percentile) and the third quartile Q3 (the 75th percentile). The middle black line of the box indicates the median (or the 50th percentile), while the length of the box (Q3 − Q1) measures the spread in the data (i.e., the interquartile range or IQR). The dashed line ("upper whisker") that extends from Q3 represents the smallest value between the maximum sample value and Q3 + 1.5 × IQR, while the dashed line ("lower whisker") that extends from Q1 represents the largest value between the minimum sample value and Q1 − 1.5 × IQR. Overall, the differences between the summary statistics are small across all cases.

**Universal characteristics of civil unrest**

In our earlier discussion, we explored the hypothesis of recurrent latent phase shifts in civil unrest, which is prompted by previous micro-level modeling studies of civil unrest (e.g., see Supplementary Methods). These studies suggest that there are universal mechanisms governing civil unrest dynamics. This raises the interesting question of whether certain elements of the macro-level phase model exhibit universality across diverse countries and regions, despite their inherent idiosyncrasies. To investigate this possibility, we analyze several features that describe the macro-level phase model and examine their behavior across countries and continents. The first two features we utilize are dimensionless quantities that capture the relationships between the phase-dependent distributions associated with the low- and high-intensity latent phases. Specifically, we focus on the relative values of the mean civil unrest magnitudes (and variances) corresponding to the high- and low-intensity latent phases. Additionally, for countries experiencing multiple phases of civil unrest, we calculate the average duration of time that a country spends in the heightened (or lowered) latent phase before transitioning to the other phases (see Methods).

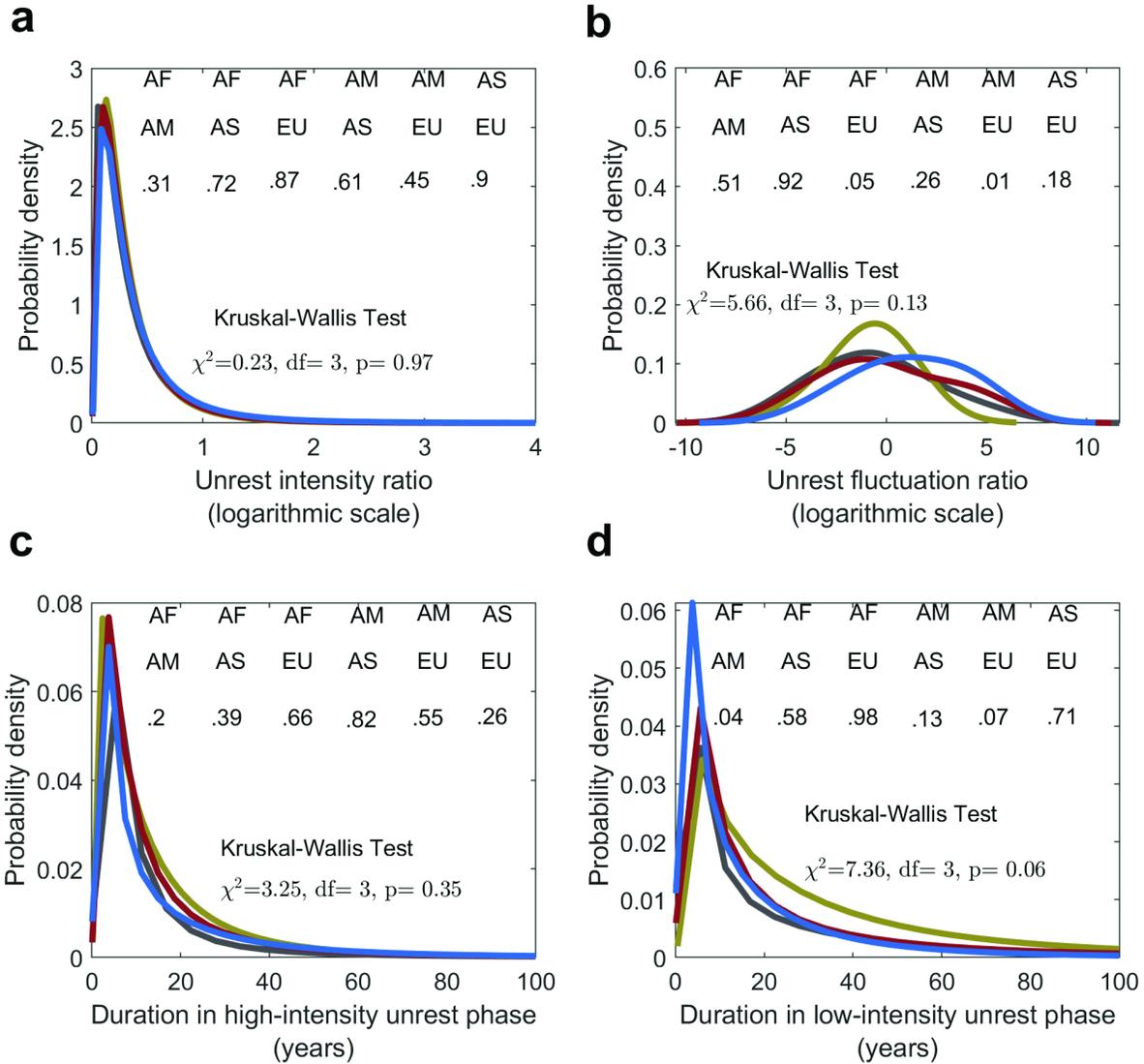

**Fig. 4. Distributions of model features by continent**. Panel **a** displays kernel density estimates of the unrest intensity ratio distributions for Africa (AF; gray solid line), America (AM; yellow-green solid line), Asia (AS; red solid line), and Europe (EU; olive-green solid line). Each country's unrest intensity ratio, defined as the ratio of the mean civil unrest magnitude during the high-intensity phase to the mean civil unrest magnitude during the low-intensity phase, is determined using the estimated macro-level phase model. The Kruskal-Wallis test's p-value of 0.97 suggests that the null hypothesis, which states that the unrest intensity ratio distributions across the four continents are identical, cannot be rejected at the 5% significance level. This finding is further supported by two-sample Kolmogorov-Smirnov pairwise tests with Bonferroni-adjusted alpha values of 0.0083 for each test. Comparable results are also observed for the unrest fluctuation ratio (Panel **b**) and the durations of high- and low-intensity phases (Panels **c** and **d**). Panel **c**,

for example, shows that the estimated length of time that a nation spends in the phase of highest intensity varies significantly, with a mode of about 4 years and possible durations exceeding 50 years.

Figure 4 displays kernel density distributions of the various model features, computed for four different continents. A Kruskal-Wallis test showed no statistically significant differences between continents for any of the evaluated features, indicating that no region tends to yield higher observed values than others. To further test the hypothesis that feature distributions are comparable across continents, we ran pairwise Kolmogorov-Smirnov tests with Bonferroni-adjusted alpha values of 0.0083 per test (0.05/6). These comparisons revealed no statistically significant differences between any of the four distributions.

The findings of our analysis are particularly intriguing because they suggest that universal distributions may govern the duration of a country's latent phases of lowest and highest intensity, despite the diversity of factors that drive civil unrest in different geographic regions. Additionally, our analysis reveals that the distributions of unrest intensity and fluctuation ratios do not significantly vary across geographic regions. These observations lend further credence to the notion that general mechanisms may underlie the dynamics of civil unrest, irrespective of the unique characteristics of nations and geographic regions.

**The geographic clustering of civil unrest**

Interactions among nations on both regional and global scales suggest that civil unrest is influenced not only by the internal dynamics of individual countries, but also by their geographic context. It is therefore important to examine whether the macro-level phase model of civil unrest exhibits any spatial clustering patterns. Such patterns can offer insights into how civil unrest in one country may be linked to neighboring countries or regions, as well as help identify the underlying causes and factors that contribute to civil unrest.

In our geographic analysis below, we use a numerical scale to evaluate a country's long-term civil unrest magnitude per unit of time. This magnitude scale of civil unrest captures the intensity of civil unrest over an extended period and enables cross-country comparisons (see Methods for more information). Figure 5 displays color-coded maps of the world's continents, depicting the magnitude scale of civil unrest in countries with plausible macro-level phase models. The data presented indicates that the magnitude scale of civil unrest varies significantly across diverse geographic regions. This observation is substantiated by the application of a Kruskal-Wallis test, which reveals a statistically significant variation in the magnitude scale of civil unrest between continents ($\chi^2(3) = 14.48$, p $=$ 0.002). This finding suggests that while the universal distributions related to various aspects of the macro-level phase model (as illustrated in Figure 4) remain unaffected by geographic location, the long-term intensity of civil unrest per unit of time may indeed be influenced by a country's geographic embedding. Taken together, these observations propose that civil unrest can arise from mechanisms that are independent of geography as well as mechanisms whose intensity varies depending on a country's unique circumstances and geographic context, including the frequency of various event types. Alongside the color-coded maps presented in Figure 5, the figure exhibits the most probable sequence of latent phases of civil unrest for representative countries from each continent. These sequences were derived by applying the Viterbi algorithm to observable historical time series data on civil unrest magnitudes, as outlined in the Methods section. The most likely sequence of latent phases of civil unrest underscores the differentiation between intermittent spikes in unrest behavior and latent phases. For instance, the peaks in political violence witnessed in Egypt during 1954, 1986, 1992, and 1993 actually constitute a phase of low-intensity, in contrast to the upsurge in late 2010. This particular escalation marked the transition into a high-intensity latent phase, characterizing the onset of the Egyptian Arab Spring. Likewise, the noticeable surge during the 2009 Iranian presidential election protests was in fact a manifestation of a low-intensity latent phase of civil unrest that extended from 1982 to 2015.

To analyze and detect the regional patterns of civil unrest suggested in Figure 5, we employed a variety of geographic statistical methods. Firstly, we conducted a random permutation test of spatial autocorrelation using Moran's I statistic (97, 98). The test revealed a statistically significant geographic correlation between a nation's own scale of civil unrest magnitude and the levels of its neighboring countries within a distance band of 100 kilometers from each country's outermost borders (Moran's I $= 0.15$, p $= 0.04$). To compile the list of neighboring countries, we also considered a contiguity spatial weight matrix, which indicates whether two countries share a border. We obtained similar results utilizing this analysis (Moran's I = 0.185, p = 0.01).

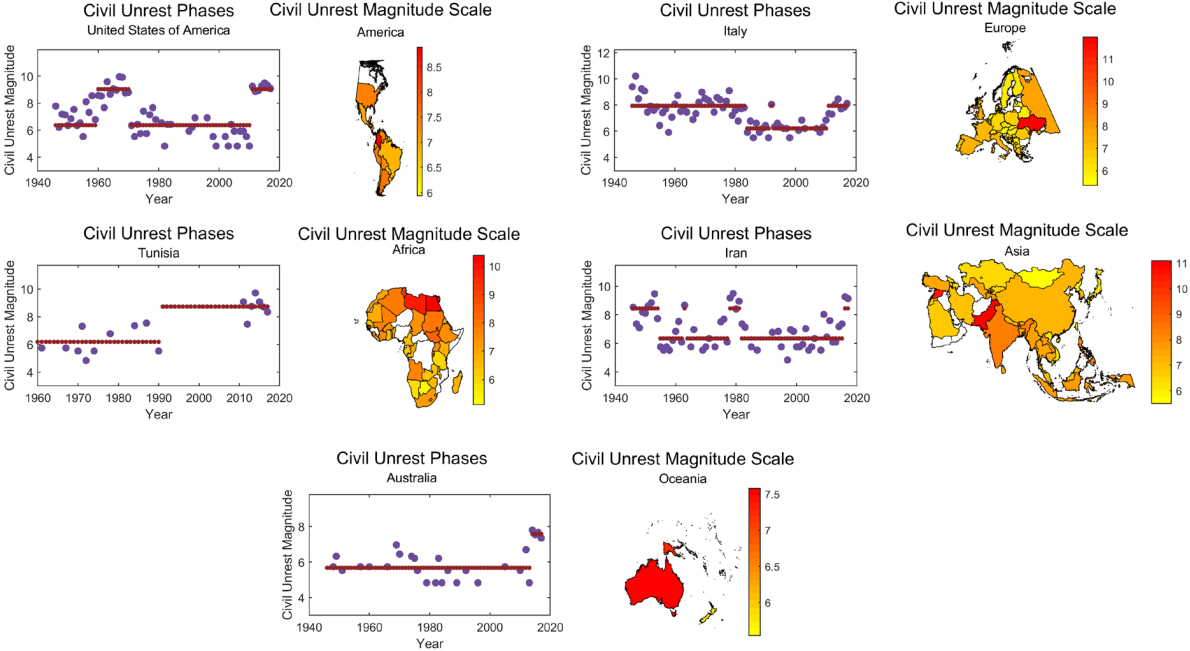

**Fig. 5. Geographical distribution of civil unrest magnitude scale around the world.** The maps display a color-coded representation of the world's continents based on the level of civil unrest magnitude scale, with excluded countries displayed in white. The magnitude scale is calculated as explained in Methods. The amplitude of the top 5% of civil unrest behavior varies between 8.259 and 8.89 in America, 7.73 and 11.99 in Europe, 9.13 and 10.39 in Africa, 9.59 and 11.09 in Asia, and between 7.2 and 7.29 in Oceania. For each representative country of a continent, we present a plot depicting the most likely sequence of civil unrest latent phases based on the historical observable dataset of civil unrest magnitudes (i.e., the logarithm of the weighted civil conflict metrics). The Viterbi global decoding algorithm was used to establish the most probable latent phase sequence (see Methods), with the phase-dependent means represented by the horizontal lines on the graphs. The most probable sequence of civil unrest latent phases highlights the distinction between spiky unrest behavior and latent phases. Country maps were generated from freely available 1:10m shapefiles sourced from Natural Earth, accessible to the public at naturalearthdata.com.

The Moran's I statistic reveals that the regional distribution of high and/or low civil unrest magnitude scales is more geographically clustered than what one would expect by chance. However, it does not capture unexpected spikes of high or low civil unrest magnitude scales. To identify such patterns, we conducted random permutation tests of spatial clustering using the Getis-Ord General G* statistic (99, 100). Our analysis found that the observed General G* was significantly higher than the expected General G*, indicating that the spatial distribution of high civil unrest magnitude scales is more geographically concentrated than what one would predict by chance (G*=0.038, p=0.047). This result provides strong evidence that there are spatial hotspots of civil unrest that cannot be explained by random spatial processes alone.

To identify the locations where large or small magnitude scales of civil unrest cluster spatially, as suggested by the Getis-Ord General G* statistic, we conducted a random permutation test of local clustering using the Getis-Ord Local $G_i^*$ statistic (99, 100). Low p-values from the test indicate statistically significant high levels of civil unrest in a country and its neighboring countries (hotspots), while high p-values indicate statistically significant low levels of civil unrest in a country and its neighbors (coldspots). As stated previously, we consider nations to be neighbors if they are within 100 kilometers of each other. Figure 6 shows a map of abnormal geographical concentrations of high or low values (hotspots or coldspots) of civil unrest magnitude scales around the globe.

The global hotspot analysis of civil unrest magnitude scale reveals a geographic cluster of countries with statistically significant high levels of civil unrest magnitude (hotspots).

These countries are primarily located in North-Central Africa (Algeria, Tunisia, Libya, Chad, Sudan, and Egypt), the Middle East (Israel, Lebanon, Syria, Jordan, and Saudi Arabia), Eastern Europe (Belarus and Moldova), South Asia (India, Pakistan, and Sri Lanka), Southeast Asia (Myanmar), and Central Asia (Tajikistan). In contrast, the majority of clusters with low values of civil unrest magnitudes (coldspots) are located in Southern Europe (Albania, Bosnia and Herzegovina, Croatia, Montenegro, North Macedonia, Republic of Serbia, Kosovo), Eastern Europe (Czech Republic), Western Europe (Austria, Belgium, Germany, Netherlands), Northern Europe (Denmark, Sweden), Eastern Asia (South Korea), South America (Suriname), Middle Africa (Angola), and Southern Africa (South Africa).

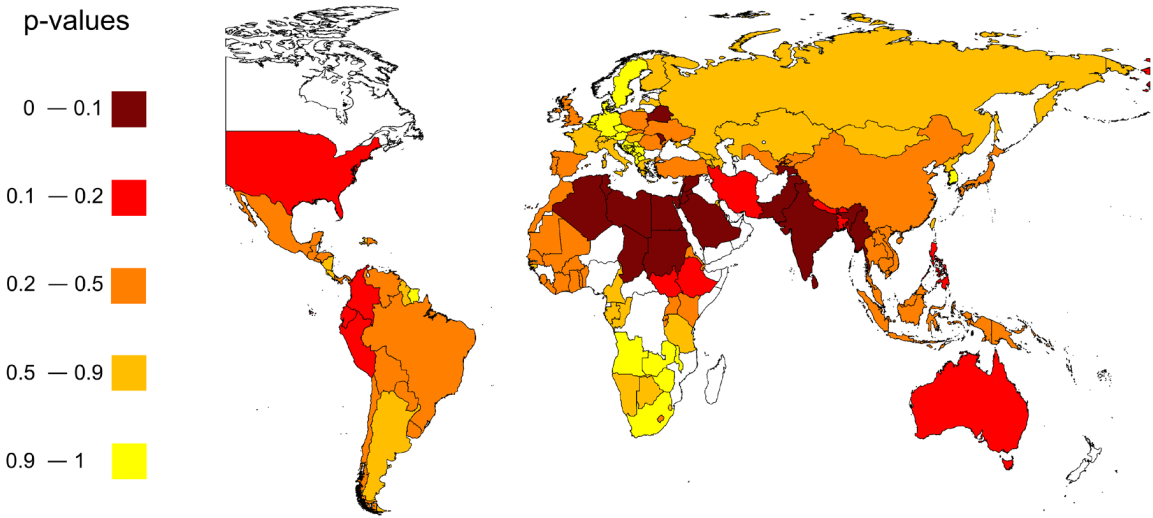

**Fig. 6. Global hotspot and coldspot analysis of the civil unrest magnitude scale.** The map displays the significance (p-value) of the local concentration of civil unrest in each country. The p-values were calculated using the Getis-Ord Local $G_i^*$ statistic and a random permutation test for local clustering.

We used a contiguity spatial weight matrix to determine whether the boundaries of two countries are within 100 kilometers of each other. The magnitude scale for civil unrest is defined in Methods. Low p-values (p-value≤0.1) indicate statistically significant levels of civil unrest in a country and its neighboring countries (hotspots), while high p-values (p-value≥0.9) indicate statistically significant low levels of civil unrest in a country and its neighbors (coldspots). Country maps were generated from freely available 1:10m shapefiles sourced from Natural Earth, accessible to the public at naturalearthdata.com.

Figures 5 and 6 demonstrate that civil unrest tends to occur in concentrated areas, similar to other geographically localized events (101). However, it is important to note that the spatial clustering analysis employed in these figures cannot differentiate between cases where the geographical clustering of civil disorder results from an "apparent contagion" and cases where the clustering results from a "true contagion" process (102). For instance, patterns of civil unrest may result from geographical contagion processes, such as cross-border influence during the Arab Spring or Eastern Europe 1989-91, or external factors like shared regional economic and political conditions. Although the spatial correlations presented in Figures 5 and 6 cannot establish causal contagion mechanisms, exploring richer datasets can shed light on spillover and social diffusion effects that may account for these correlations.

Importantly, the variable examined in Figures 5 and 6 is derived directly from the macro-level phase model. Specifically, the civil unrest magnitude scale is computed based on the equilibrium distribution of the Markov chain governing the transitions between the latent civil unrest phases and their expected magnitudes (see Methods). Furthermore, despite separately estimating the macro-level phase model parameters for each country, the collective hotspot pattern depicted in Figure 6 aligns with historical records of civil unrest worldwide. A notable example is the Arab Spring, characterized by a series of protests, uprisings, and revolutions that began in late 2010 and continued into the early 2010s across several countries in the Middle East and North Africa region. Key countries associated with the Arab Spring include Tunisia, Libya, Egypt, and Syria. Ongoing uprisings and protests in Algeria, Sudan, Lebanon, and Egypt since late 2018 have been seen as a continuation of the Arab Spring. The persistent conflicts in Syria, Libya, and Lebanon are viewed as aftermaths of the Arab Spring, resulting in political instability, economic hardships, and crises. The hotspot pattern illustrated in Figure 6 suggests that, although each country had its distinctive context and specific triggers for the protests, several shared factors contributed to the broader wave of uprisings and movements associated with the Arab Spring. Consequently, the observed geographical clustering of civil unrest in the figures, an emergent characteristic of the macro-level phase model, aligns with historical records of civil unrest on a global scale. This alignment reinforces the idea that recurring latent phases are a fundamental feature of civil unrest behavior.

## DISCUSSION

The study of complex systems generally entails exploring the underlying laws that govern the phase transitions of self-organized phenomena, as observed in a wide range of complex systems (e.g., 1, 3, 82). In this study, we analyze a large dataset of civil unrest events from 1946 to 2017 in 170 countries to investigate the phenomenology of phase transitions that underpin the severity of collective civil unrest. History has shown that civil unrest is characterized by a wide range of intensities, and there is no such thing as a "normal" level of violence or discontent expression. Although there has been a notable surge in high-intensity anti-government demonstrations in the past two decades, the majority of civil unrest events are of low to moderate intensity. To shed light on this seemingly unpredictable pattern of phase transitions, our study introduces a macro-level statistical model of civil unrest, which enables us to better understand and quantify this behavior. The macro-level phase model posits that collective civil unrest occurs in recurring, discontinuous latent phases that can be detected and quantified. To capture this phenomenon, the phase model utilizes an underlying Markov chain that governs the transitions between different latent phases, with the current latent phase determining the magnitude of civil unrest at any given time. The primary finding of this study is that the dynamics of collective civil unrest can be considered a discontinuous process characterized by unique, measurable, repeating latent phases. The macro-level phase model statistically represents the transitions between these latent phases as a Markov process, where moments of relative peace alternate with large-scale civil unrest. These latent phase changes may stem from various phenomenological origins. Economic disparity, governmental instability, social injustice, and ethnic or religious tensions, among other factors, can contribute to tensions building up during the low-intensity latent phase of civil unrest (32, 37, 47, 51). Individuals are more likely to take action and demand change when societal and political tensions reach a tipping point. However, when tensions are low or there are no specific triggering events, the transition to the high-intensity latent phase of civil unrest may be less likely. It is important to note that a triggering event or series of events, such as contentious election results or high-profile acts of police brutality, may appear as a sudden and intense burst of civil unrest activity but may not necessarily indicate a transition to or the presence of a high-intensity latent phase. In other words, spiky behavior in observable civil unrest time-series data may be the result of noise that could mask the undetected latent phase of civil unrest if not properly understood.

The approach outlined in this paper for detecting phase changes in civil unrest remains effective, even in cases where a comprehensive understanding of all the mechanisms driving civil unrest is lacking. However, to establish theoretical foundations for our empirical research and enhance its credibility, we grounded the primary hypothesis (and subsequently the macro-level phase model) of this paper—civil unrest following a discontinuous phased pattern—on models that view civil unrest as a self-organized critical phenomenon (52, 58, 76—79; see example in Supplementary Methods). Instead of providing an exact representation of civil unrest mechanisms, these models simulate the gradual accumulation of tensions or vulnerabilities during a relatively peaceful period. These tensions are then released through sudden transitions into widespread civil unrest, offering plausible insights into the possibility of recurrent critical shifts in real-world civil unrest data. This aspect necessitates additional empirical investigation, which is the current focus of our work. The micro-dynamic spatial mechanisms in this class of self-organized critical models are founded on the concept that groups of individuals, connected via a spatial network of closely interlinked geographic regions, partake in collective civil unrest owing to shared long-term social, economic, and political stress, as well as peer influence (52, 70—73). The progression of civil unrest within this network of affected individuals and locations may then exhibit a discontinuous phased pattern as a plausible outcome of these micro-dynamic mechanisms (see Supplementary Methods). More precisely, a threshold effect in the form of slowly changing intrinsic parameters—such as the count of regions susceptible to social, economic, and political stress—may give rise to the discontinuous nature of civil unrest and the accompanying latent phases. Once a specific threshold is exceeded, civil disturbance activities are prone to extensive expansion, rapidly escalating the level of civil disorder. This escalation leads to the widespread dissemination of civil unrest across social networks and geographic regions. In this case, we anticipate a shift in the dynamics of collective civil unrest from small-scale, low-intensity civil disorder to large-scale, high-intensity civil disorder. The primary objective of this study was to examine whether the hypothesized latent phases of civil unrest, as anticipated by such theoretical models, correspond to actual civil unrest events across various countries and time periods. To accomplish this objective, we conducted a comprehensive evaluation of the macro-level phase model using large-scale civil unrest data, maximum-likelihood estimation, and Monte Carlo Kolmogorov-Smirnov two-sample testing (see Methods). Our findings provide evidence supporting the null hypothesis that the macro-level phase model and its proposed discontinuous nature of civil unrest accurately reflect the characteristics of civil unrest data from various countries worldwide.

After establishing the statistical plausibility of the macro-level phase model, we proceeded to examine its consistency with the presence of a significant positive spatial correlation between a country's own levels of civil unrest and those of its neighbors. This correlation is anticipated by both the historical account of civil unrest events and the spatial localized mechanisms (i.e., the rapid spread of civil unrest in one region to neighboring regions) employed in self-organized critical models of civil unrest (52, see Supplementary Methods). To assess the consistency of the macro-scale phase model with the geographical clustering effect of civil disorder, we utilized various spatial statistical analysis methods that incorporated a numerical scale determined directly by the phase model's parameters. This scale quantified the magnitude of a country's long-term unrest per unit of time. Our findings revealed that this civil unrest magnitude scale is correlated across neighboring countries, indicating that civil unrest events worldwide are spatially clustered. Furthermore, further analysis demonstrated that the magnitude of civil unrest is concentrated in specific geographic regions. The observed geographical clustering of civil unrest can be attributed to either 'apparent contagion' or 'true contagion' processes (102). 'Apparent contagion' arises when countries sharing socio-cultural backgrounds are prone to similar civil unrest due to similarities in grievances and expressions. It can also occur among countries with strong economic or political ties, given the influence of these factors on protests and uprisings. On the other hand, 'true contagion' involves countries influencing each other's civil unrest through communication and social networks, facilitating the spread of unrest between nations. Extensive media coverage of protests in one country, for instance, can inspire similar events in others. A prospective future study could focus on investigating whether contagion processes ('true contagion') play a role in explaining the global dissemination of civil unrest, leading to geographic clustering. One approach involves detecting hotspots and coldspots of civil unrest through a contiguity matrix, wherein nations are deemed 'neighbors' if they share comparable socio-cultural variables, without regard for geographical proximity. If our analysis reveals clusters that encompass countries not contiguous in geography, it implies that socio-cultural similarities are the probable driving force behind the clustering. On the contrary, if we identify a cluster primarily based on geographical proximity that doesn't align with socio-cultural similarities, it suggests the presence of a contagion process influencing the geographic clustering.

The methodology presented in this paper has a dual purpose: it can validate theoretical models against empirical findings and independently detect potential critical transitions, even when a thorough understanding of all the pertinent mechanisms underlying real-world civil unrest is lacking. This latter facet offers several promising applications and policy implications.

A key application of the macro-level phase model lies in forecasting. For example, estimating the parameters of the macro-level phase model (as detailed in Supplementary Figures and Tables) allows for immediate computation of the conditional distribution of observed civil unrest magnitudes for a specified forecast horizon. Moreover, the model facilitates 'phase prediction,' allowing for the determination of the conditional distribution of the phase for a future horizon. This goes beyond the application of the Viterbi algorithm (as described in Methods) used to establish the most probable sequence of latent phases of civil unrest for past and present phases. Additionally, forecasting can be accomplished by integrating the macro-level phase model with classification-based methodologies, such as multinomial logistic regression. After determining the most probable sequence of latent civil unrest phases (as described in Methods), multinomial logistic regression can be utilized to predict transition probabilities between these phases based on a set of independent factors (such as country-specific socio-economic data). Continuous monitoring of the parameters of the macro-level phase model using high-resolution and comprehensive ongoing unrest data, and utilizing the aforementioned forecasting techniques, could operate as an early warning system for heightened susceptibility to social instability.

Our methodology for detecting phase transitions is closely aligned with research on early-warning signals for critical transitions in highly complex systems (1, 2, 20, 24). Drawing from simple models of catastrophic bifurcations and analyses of simulation models subject to stochastic forcing, this research suggested the existence of generic indicators of impending transitions that can manifest in highly complex systems, even without a complete understanding of all the underlying mechanisms (2). Specifically, it was posited that as the system gradually approaches a catastrophic bifurcation, there is a noticeable increase in lag-1 autocorrelation (1, 2), heightened variance (1, 2, 20, 24), or other relevant indicators (1, 2). Similar signals may appear as the system approaches a critical threshold that is not associated with catastrophic bifurcations (1, 2). Furthermore, these indicators may appear not only in basic models but also in complex, detailed models that closely represent spatially intricate systems (1, 2). An intriguing aspect of our research involves employing the macro-level phase model to investigate the resilience of these signals within the realm of civil unrest dynamics. One particular objective is to ascertain if there is a discernible increase in autocorrelation or variance as society steadily approaches a phase transition (as independently detected through the Viterbi algorithm, see Methods). Analyzing civil unrest time-series collected on shorter time scales (days instead of years) could provide insights into this question. Finally, as pointed out in (2), the challenge of early-warning signal detection in real data introduces a significant hurdle, potentially yielding both false positives and false negatives. For instance, it might fail to identify early-warning signals even when a sudden transition has indeed taken place. In such scenarios, (2) suggested the need for a robust set of statistical procedures to ascertain the significance of an increase in autocorrelation or variance. Our approach to detecting and quantifying phase transitions, based on a latent statistical model, demonstrates reduced sensitivity to data fluctuations. Consequently, it can effectively address and complement the limitations associated with the earlier mentioned early-warning signal methodologies.

One of the primary challenges in practical policy implementation lies in the capacity to anticipate phase transitions of civil unrest in advance, enabling timely initiation of preventive measures and adequate preparations before the transition takes place. The discussed forecasting techniques could be valuable in this regard. From the perspective of government and relevant authorities engaged in risk management, they can formulate strategies to effectively mitigate the risk of undesired transitions. Mechanistic models, like the one outlined in Supplementary Methods, could provide insights for organizing social systems to reduce vulnerability to widespread, high-intensity civil unrest. An effective strategy to attain this objective involves creating an environment that fosters the resolution of social, economic, and political grievances prevalent among different segments of the population. Disregarding the root causes that nourish these resentments and being deceived by a prolonged period of apparent calm over several years could significantly heighten these grievances to a critical point. In such a state, a minor spark could easily ignite a widespread conflagration. Several international organizations, including the United Nations (UN) and the Organization for Security and Co-operation in Europe (OSCE), focus on monitoring civil unrest and related concerns. These organizations could make use of the indicators outlined in this paper, including the magnitude scale of civil unrest (refer to Figures 5-6), to assess and categorize countries and regions on a spectrum from vulnerability to resilience. Such evaluations can subsequently guide policy development and decision-making. Finally, from the viewpoint of civil society, identifying opportunities to facilitate desired phase transitions in social systems, especially under totalitarian regimes, can hold significant value. One strategy involves establishing interconnected social or communication networks across diverse regions within a country, which could trigger widespread unrest and subsequent phase shifts.

The shift from low-intensity to high-intensity civil unrest is a critical sign of social instability within a country, and throughout history, this transition has often coincided with rapid societal changes (27, 103). As a result, comprehending the emergence of collective civil unrest, which can manifest abruptly and unpredictably, requires the creation of efficient early warning systems that can detect and forecast phase transitions of civil disorder across nations and throughout time. The identifiable macroscopic latent phases of social unrest likely arise as a result of accumulating tensions and their rapid release through spatial networks that link closely associated individuals or regions within the country. Understanding the factors that contribute to these spatial mechanisms, as well as employing methods to detect phase transitions, can provide valuable insights for devising strategies aimed at mitigating the impact of civil unrest. Finally, our approach for detecting and measuring phase transitions in collective civil unrest have broad applications beyond its immediate scope and can be utilized to analyze various other collective human phenomena.

## METHODS

**Domestic conflict event data**

This research utilizes a long-term dataset from the Cross National Time Series Dataset (86) to track civil unrest events in 170 countries from 1946 to 2017. The New York Times is the primary source of information on civil disorder. Assassinations, general strikes, guerrilla warfare, major government crises, purges, riots, revolutions, and anti-government demonstrations are the eight domestic conflict event types analyzed (86, 104). Accordingly, the weighted conflict (WC) metric was computed for each nation and each year as follows (86, 104):

$$\text{WC} = \left(\frac{25e_1 + 20e_2 + 100e_3 + 20e_4 + 20e_5 + 25e_6 + 150e_7 + 10e_8}{8}\right)100 \quad (1)$$

where $e_i$ represents the frequency of event type $i$. The magnitude of civil unrest was then determined as the logarithm of the weighted conflict metric.

**Basic definitions of the macro-level phase model**

The civil unrest phase model is described in terms of two stochastic processes: a stochastic process of unobserved latent phases $\{Q_t, t = 1,2,...\}$ where $Q_t$ denotes the unobserved latent phase at time $t$; and a stochastic process of observed civil unrest magnitudes $\{X_t, t = 1,2,...\}$. Here, each $Q_t$ can take values in the set $\{1,2,...,m\}$, where $m$ is the number of unobserved latent phases, and each $X_t$ can take any real value. The dynamics of the model is captured by the following equations:

$$P(Q_t|\mathbf{X}^{(t-1)}, \mathbf{Q}^{(t-1)}) = P(Q_t|Q_{t-1}), \quad t = 2, 3, ... \quad (2)$$

$$P(X_t|\mathbf{X}^{(t-1)}, \mathbf{Q}^{(t)}) = P(X_t|Q_t), \quad t \in \mathbb{N} \quad (3)$$

Let $a_{ij}$ denote the one-step transition probability $P(Q_t = j|Q_{t-1} = i)$, and let $\mathbf{A}$ be the matrix of one-step transition probabilities $a_{ij}$. That is, we assume that the stochastic process underlying the dynamics of the unobserved civil unrest latent phases is a Markov chain. We suppose that the phase-dependent probability density function of $X_t$ in latent phase $i$, $P(X_t|Q_t = i)$ is normal with unobserved parameters $\mu_i$ and $\sigma_i^2$. Equivalently, if $Y_t$ is the weighted conflict metric given latent phase $Q_t = i$, then $Y_t = e^{X_t}$ has a log-normal distribution. For all countries, we refer to $i = 1$ for which $\mu_1 = \max_{j \in \{1,2,\ldots,m\}} \mu_j$ as the highest-intensity latent phase, and $i = m$ for which $\mu_m = \min_{j \in \{1,2,\ldots,m\}} \mu_j$ as the lowest-intensity latent phase. Finally, we denote the initial distribution of the Markov chain by $\boldsymbol{\rho} = (\rho_1, \rho_2, \ldots, \rho_m)$.

**Marginal distributions of observed civil unrest magnitudes**

Let $p_i(x) \equiv P(X_t = x|Q_t = i)$ and let $\mathbf{P}(x)$ denote the diagonal matrix with the $i$th diagonal element $p_i(x)$. The marginal probability density function of $X_t$ is then given by $P(X_t = x) = \boldsymbol{\rho} \boldsymbol{A}^{t-1} \mathbf{P}(x) \mathbf{1}'$ where $\mathbf{1}'$ is a column vector of ones. Averaging the values of $P(X_t = x)$ throughout all years yields the marginal probability density function of observed civil unrest magnitudes at any arbitrary time point. Monte Carlo simulations are another method for calculating the marginal density function of observed civil unrest magnitudes. Specifically, we produce synthetic data of simulated civil unrest magnitudes using the estimated macro-level phase model, and then use the synthetic data to obtain the kernel density estimate. This technique is repeated with a new set of synthetic data to obtain an additional kernel density estimate. Over 5000 kernel density estimate samples were calculated in total. The marginal density function of simulated magnitudes of civil unrest was calculated by averaging the samples. Both methods were found to yield similar results. In this paper (see Figure 3), the marginal cumulative distribution functions (CDFs) of civil unrest magnitudes as predicted by the macro-level phase model were obtained using the Monte Carlo method.

**Parameter estimation of the macro-level phase model**

The macro-level phase model has altogether $m(m+2)$ parameters, which are estimated by maximizing the likelihood of the observed civil unrest magnitude data. The likelihood function is the joint density of the data, and is given by the following:

$$\mathcal{L}_T(\theta) = \boldsymbol{\rho}\mathbf{P}(x_1)\mathbf{AP}(x_2)\mathbf{AP}(x_3)\cdots\mathbf{AP}(x_T)\mathbf{1}' \tag{4}$$

Where $\theta = \{\rho_i, a_{ij}, \mu_i, \sigma_i^2 | i,j = 1,2,\ldots,m\}$ is the set of model's parameters, $T$ is the length of the civil unrest time series, and $\mathbf{x}^{(T)} = (x_1, x_2, \ldots, x_T)$ is the vector of observed civil unrest magnitudes. The log-likelihood function is defined by $\ell_T(\theta) = log\mathcal{L}_T(\theta)$. The Baum-Welch algorithm was used in this paper to get the maximum likelihood estimates of the parameters (87—92). The Baum-Welch algorithm is an iterative procedure that begins with parameter estimates. The parameters are iteratively adjusted until convergence is obtained. Although it is outside the scope of this paper, we discuss the algorithm's major steps briefly (see 87—92 for more information). The algorithm includes two quantities known as forward and backward probabilities, which are as follows:

$$\boldsymbol{\alpha}_t = \boldsymbol{\nu}\mathbf{P}(x_1)\prod_{k=2}^{t}\mathbf{AP}(x_k) \tag{5}$$

$$\boldsymbol{\beta}'_t = \left(\prod_{k=t+1}^{T}\mathbf{AP}(x_k)\right)\mathbf{1}', \quad t = 1,2,\ldots,T-1 \tag{6}$$

and $\boldsymbol{\beta}'_T = \mathbf{1}'$. It can be shown that $\alpha_t(i) = P(\mathbf{X}^{(t)} = \mathbf{x}^{(t)}, Q_t = i)$ and $\beta_t(i) = P(X_{t+1} = x_{t+1}, X_{t+2} = x_{t+2}, \ldots, X_T = x_T | Q_t = i)$. We also need the two following definitions:

$$u_i(t) = P(Q_t = i | \mathbf{x}^{(T)}) = \alpha_t(i)\beta_t(i)/\boldsymbol{\alpha}_t \boldsymbol{\beta}'_t \tag{7}$$

$$v_{ij}(t) = P(Q_{t-1} = i, Q_t = j | \mathbf{x}^{(T)}) = \alpha_{t-1}(i)a_{ij}p_j(x_t)\beta_t(j)/\boldsymbol{\alpha}_t \boldsymbol{\beta}'_t \tag{8}$$

Having defined the above probabilities, the Baum–Welch iterative estimation equations for $\rho_i$, $a_{ij}$, $\mu_i$, and $\sigma_i^2$ are calculated as follows, based on the current parameter estimates:

$$\hat{\rho}_i = u_i(1) \tag{9}$$

$$\hat{a}_{ij} = \sum_{t=2}^{T} v_{ij}(t) \bigg/ \sum_{j=1}^{m} \sum_{t=2}^{T} v_{ij}(t) \tag{10}$$

$$\hat{\mu}_i = \sum_{t=1}^{T} u_i(t) x_t \bigg/ \sum_{t=1}^{T} u_i(t) \tag{11}$$

$$\hat{\sigma}_i^2 = \sum_{t=1}^{T} u_i(t)(x_t - \mu_i)^2 \bigg/ \sum_{t=1}^{T} u_i(t) \tag{12}$$

**Macro-level phase model selection**

We studied macro-level phase models with varying numbers of latent phases. There are several approaches to determining the appropriate number of latent phases. We use the Bayesian information criterion (BIC), which is defined as follows (92):

$$\text{BIC} = -2\ell_T(\theta) + n \log T \tag{13}$$

where $\ell_T(\theta)$ is the estimated model's log-likelihood, $n = m(m + 2)$ is the number of model parameters, and $T$ is the number of civil unrest observations.

**The treatment of zero observations**

The magnitude of civil unrest is represented as the logarithm of the weighted civil conflict metric, making it difficult to directly apply the macro-level phase model to any civil unrest time series with zero weighted civil conflict values. However, the domestic conflict event data includes zero event counts, and non-zero weighted civil conflict values appear to be a combination of lognormal distributions. It is important to note that the zero event counts are likely due to the geographic bias and limited scope of the newspaper reports from which the event data is derived. In this case, zero-event data can be viewed as undetected civil unrest events with measured values below the detection limit and therefore unavailable for statistical analysis. To address this, it may be appropriate to use a conservative approach that considers zero-event observations as missing values rather than more sophisticated approaches, such as adding a positive constant to all sample values or using more general phase-dependent distributions like the zero-modified lognormal/normal distributions.

To account for missing values in the Baum-Welch algorithm, we simply substitute the diagonal matrices $\mathbf{P}(x_t)$ corresponding to missing observations $x_t$ with the identity matrix (105).

**Phase model metrics: civil unrest magnitude scale and phase durations**

Let $\boldsymbol{\pi} = (\pi_1, \pi_2, \ldots, \pi_m)$ represent the limiting probabilities associated with the transition probability matrix $\mathbf{A}$. Thus, $\pi_i$ equals the fraction of time a country will spend in latent phase $i$ in the long term. The civil unrest magnitude scale is defined as the long-run mean unrest magnitude per unit of time, provided by $\sum_{i=1}^{m} \pi_i \mu_i$ where $\mu_i$ are the estimated parameters corresponding to the phase-dependent normal distributions. The average length of time a country spends in the lowest and highest intensity phases can be calculated as follows: $\bar{L} = 1/\sum_{j \neq m} a_{mj}$ and $\bar{H} = 1/\sum_{j \neq 1} a_{1j}$.

**Global decoding and the Viterbi algorithm**

The estimated parameters of the macro-level phase model are used to determine the most probable sequence of latent phases, which is achieved by finding the sequence of latent phases $\mathbf{q}^{(T)} = \{q_1, q_2, \ldots, q_T\}$ that has the highest conditional probability $P\bigl(\mathbf{Q}^{(T)} = \mathbf{q}^{(T)} | \mathbf{X}^{(T)} = \mathbf{x}^{(T)}\bigr)$. This problem, known as global decoding, is solved by utilizing the dynamic programming algorithm of Viterbi (90, 106). Figure 5 illustrates the predicted latent phases of civil unrest for representative countries, as determined by the Viterbi algorithm.

**Assessing the goodness of fit of the macro-level phase model**

After estimating the parameters of the civil unrest phase model through the Baum-Welch maximum likelihood estimation procedure, we employed Monte Carlo Kolmogorov-Smirnov two-sample tests to evaluate the plausibility of the macro-level phase model. The analysis was conducted using the observed civil unrest dataset, which comprised data from 170 countries from 1946 to 2017. Our aim was to demonstrate that the macro-level phase model is reasonable for many countries and cannot be rejected with a high level of confidence.

The Monte Carlo goodness-of-fit test is based on comparing the marginal distribution generated by the macro-level phase model to the empirical distribution of civil unrest data. In this study, we employed the Kolmogorov-Smirnov (KS) statistic, which measures the largest vertical difference between the empirical and postulated distribution functions. To perform the test, we first estimated the parameters of the macro-level phase model using the Baum-Welch algorithm and real unrest data. We then compared the empirical distribution function of the civil unrest data with the marginal distribution obtained by the estimated phase model to calculate the empirical KS statistic. Next, we generated synthetic data based on the estimated macro-level phase model and estimated a new macro-level phase model based on the generated synthetic data. This process was repeated multiple times with additional sets of synthetic data, resulting in over 400 samples of KS statistics in total. We computed the p-value as the proportion of simulated statistics that were greater than the empirical KS statistic. If the computed p-value was less than 0.05, we rejected the null hypothesis that the macro-level phase model adequately describes the actual unrest data. Our use of the Monte Carlo method in conjunction with the KS statistic provides a robust and reliable assessment of the core hypothesis that there are recurring latent phase shifts in civil unrest across diverse nations and time periods.

## DATA AVAILABILITY

Data on civil unrest episodes are available from Databanks International's Cross-National Time-Series Data Archive (CNTS, ISSN 2412-8082) for every country from 1946 through 2017. See https://www.cntsdata.com.

## CODE AVAILABILITY

The code used to generate results shown in this study is available from the author upon request.

## AUTHOR CONTRIBUTIONS

DB conceived and designed the research, devised the modeling methodology, curated the data, conducted the analysis, validated the results, contributed to materials and analysis tools, wrote the paper, and designed and wrote the software employed in the analysis.

## COMPETING INTERESTS

The Authors declare no Competing Financial or Non-Financial Interests.